\documentclass[12pt,preprint]{aastex}
\usepackage{graphicx,amsmath,amssymb}
\usepackage{amsmath}
\usepackage{pdflscape}
\usepackage{url}

\def\au{{\rm AU}}

\def\muas{\mu{\rm as}}

\def\rel{{\rm rel}}

\def\e{{\rm E}}

\begin{document}
\title{UKIRT-2017-BLG-001Lb: A giant planet detected through the dust}

\author{
Y.~Shvartzvald\altaffilmark{1,a},
S.~Calchi~Novati\altaffilmark{2},
B.~S.~Gaudi\altaffilmark{3},
G.~Bryden\altaffilmark{1},
D.~M.~Nataf\altaffilmark{4},
M.~T.~Penny\altaffilmark{3},
C.~Beichman\altaffilmark{5},
C.~B.~Henderson\altaffilmark{5},
S.~Jacklin\altaffilmark{6},
E.~F.~Schlafly\altaffilmark{7,b},
M.~J.~Huston\altaffilmark{3}
}
\altaffiltext{1}{Jet Propulsion Laboratory, California Institute of Technology, 4800 Oak Grove Drive, Pasadena, CA 91109, USA; yossi.shvartzvald@jpl.nasa.gov}
\altaffiltext{2}{IPAC, Mail Code 100-22, Caltech, 1200 E. California Blvd., Pasadena, CA 91125, USA}
\altaffiltext{3}{Department of Astronomy, Ohio State University, 140 W. 18th Ave., Columbus, OH  43210, USA}
\altaffiltext{4}{Center for Astrophysical Sciences and Department of Physics and Astronomy, The Johns Hopkins University, Baltimore, MD 21218, USA}
\altaffiltext{5}{IPAC/NExScI, Mail Code 100-22, Caltech, 1200 E. California Blvd., Pasadena, CA 91125, USA}
\altaffiltext{6}{Vanderbilt University, Department of Physics \& Astronomy, Nashville, TN 37235, USA}
\altaffiltext{7}{Lawrence Berkeley National Laboratory, One Cyclotron Road, Berkeley, CA 94720, USA}
\altaffiltext{a}{NASA Postdoctoral Program Fellow}
\altaffiltext{b}{Hubble Fellow}

\begin{abstract}
We report the discovery of a giant planet in event UKIRT-2017-BLG-001, detected by the UKIRT microlensing survey.
The mass ratio between the planet and its host is $q=1.50_{-0.14}^{+0.17}\times10^{-3}$, about 1.5 times the Jupiter/Sun mass ratio.
The event lies 0.35$^{\circ}$ from the Galactic center and suffers from high extinction of $A_K=1.68$.
Therefore, it could be detected only by a near-infrared survey.
The field also suffers from large spatial differential extinction, which makes it difficult to estimate the source properties required to derive the angular Einstein radius.
Nevertheless, we find evidence suggesting that the source is located in the far disk.
If correct, this would be the first source star of a microlensing event to be identified as belonging to the far disk.
We estimate the lens mass and distance using a Bayesian analysis to find that the planet's mass is $1.28^{+0.37}_{-0.44}\,M_{J}$,
and it orbits a $0.81^{+0.21}_{-0.27}\,M_{\odot}$ star at an instantaneous projected separation of $4.18^{+0.96}_{-0.88}$ AU. The system is at a distance of $6.3^{+1.6}_{-2.1}$ kpc, and so likely resides in the Galactic bulge.
In addition, we find a non-standard extinction curve in this field, in agreement with previous results toward high-extinction fields near the Galactic center.
\end{abstract}


{\section{{Introduction}
\label{sec:intro}}

Gravitational microlensing is unique in its ability to probe relatively untapped reservoirs of exoplanet parameter space \citep{Gaudi.2012.A}.
These include planets at all masses near the ``snowline'' where gas and ice giants are likely to form \citep{Ida.2005.A}, 
hence probing the frequency and mass function of snowline planets \citep{Cassan.2012.A,Shvartzvald.2016.A,Suzuki.2016.A};
planetary systems around all types of stars throughout the Galaxy, thus allowing measurement of the Galactic distribution of planets \citep{Calchi.2015.A,Penny.2016.A,Zhu.2017.A}; and free-floating planets \citep{Sumi.2011.A,Mroz.2017.A}.
However, currently only 51 planetary systems hosting 53 planets\footnote{\tt https://exoplanetarchive.ipac.caltech.edu} have been discovered using microlensing.
While new ground-based microlensing surveys with global networks of telescopes and high cadences (e.g., KMTNet; \citealt{Kim.2016.A}) are expected to detect a few tens of planets per year \citep{Henderson.2014.B}, it is clear that a space-based survey is required in order to significantly increase the number of detected systems and to detect planets with masses substantially less than that of the Earth, thus fully exploiting the microlensing potential \citep{Bennett.2002.A,Penny.2013.A}.

The proposed Wide Field InfraRed Survey Telescope ({\it WFIRST}) flagship mission \citep{Spergel.2015arXiv.A}, which is planned to launch in
mid-2020s, would dedicate $\sim$25\% of its lifetime to a microlensing survey. This survey is predicted to discover
thousands of exoplanets near or beyond the snowline via their microlensing light curve signatures (Penny et al., in preparation), enabling a {\it Kepler}-like statistical analysis
of planets $\sim$1--10 AU from their host stars and potentially revolutionizing our understanding of planet formation.
In addition to the superb photometry, high cadence, and continuous observations, the survey will be conducted in the near-infrared (NIR).
An NIR microlensing survey suffers from less extinction than traditional optical surveys, enabling observations closer to the Galactic plane and center, where 
the stellar surface density of sources and lenses, and thus the microlensing event rate, is highest.
However, until recently no dedicated NIR microlensing survey has been conducted, and so the event rate in the NIR has not been measured,
which is crucial for {\it WFIRST} field optimization \citep{Yee.2014arXiv.A}.

During the 2015 and 2016 seasons, we conducted the first NIR microlensing survey, with UKIRT, as support for the $Spitzer$ and $Kepler$ microlensing campaigns.
From examination of 5$\%$ of the 2016 UKIRT fields (overlapping with optical survey fields), we discovered five highly extinguished, low-Galactic latitude microlensing events \citep{Shvartzvald.2017.A}.
These events were not detected by optical surveys, likely due to the high extinction. Combining these detections
with additional events that were also detected by optical surveys in these fields, we found evidence that the event rate is indeed higher in this region, closer to the Galactic plane.
In 2017, we initiated a full NIR microlensing survey with UKIRT, covering all potential {\it WFIRST} fields including the Galactic plane and center (see Figure~\ref{fig:fields}), which are inaccessible to optical surveys due to the high extinction. The fields are observed with a daily cadence, allowing us to easily detect microlensing events with their typical timescale of approximately 20 days \citep{Shvartzvald.2016.A}, and with a cadence of 3 epochs/night in the central fields, which are expected to have an excess of short-timescale bulge-bulge events \citep{Gould.1995.A}.

Bound microlensing planets are discovered through their anomalous signature on the otherwise smooth single-lens light curve.
These anomalies occur when the source passes near or over a caustic. There are three types of caustics (for a thorough review see \citealt{Gaudi.2012.A}): central caustics (associated with the host), planetary caustics (associated with the planet), and resonant caustics, which occur when the central and planetary caustics merge into a single caustic. The location and shape of the caustics are determined by the mass ratio $q$ between the host and the planet, and the scaled (with respect to the Einstein radius $\theta_{\rm E}$) instantaneous projected separation $s$. 
The size of the caustics is mostly determined by the mass ratio $q$ (as the separation $s$ changes by only a factor 2 in the ``lensing zone'' \citep{Griest.1998.A}), and thus for events due to a star and a planet (and thus small $q$) the caustics are usually small. Anomalies due to planets typically last $\lesssim1$ day for planetary or central caustics, and thus require high cadence in order for them to be detected and well characterized. However, resonant caustics are always larger than either the planetary or central caustics (for a given mass ratio $q$) and thus have a larger cross section \citep{Dominik.1999.A}, as well as longer anomaly duration, and therefore can be detected even with a daily cadence.  
In fact, about 40\% of the microlensing planetary systems to date have been detected through resonant caustic perturbations\footnote{We also consider events with degenerate solutions when at least one of the solutions is due to a resonant caustic.}.
These include the first microlensing planet \citep{Bond.2004.A}, the first planet simultaneously observed from ground and space \citep{Udalski.2015.A},
and the two multiple-planet systems \citep{Gaudi.2008.A,Han.2013.A}. All of these events also had long timescales (60--150 days), thus the anomalies lasted several days.

In this Letter, we report the discovery of the planetary event UKIRT-2017-BLG-001. This is the first planet to be detected solely by the UKIRT survey.
The planetary perturbation is caused by a resonant caustic of a giant planet ($q\sim10^{-3}$) and the event has a long timescale of $\sim$100 days. Thus, the survey cadence was sufficient for the detection and characterization of the planet. The event lies close to the Galactic center (0.35$^{\circ}$) and 
in a field with significant total and spatially variable extinction, thus it could not be detected by optical microlensing surveys.

We describe the UKIRT observations and event detection in Section \ref{sec:obs}. In Section \ref{sec:model} we present
the best-fit microlensing model of the event. 
In Section \ref{sec:cmd} we derive the source properties by analyzing the color-magnitude diagram (CMD) of the event, finding indications that suggest it is a far-disk source.
In Section \ref{sec:phys} we estimate the physical properties of the planetary system using a Bayesian analysis.
An analysis of the multi-band extinction toward this field is presented in Section \ref{sec:ext_law}.
Finally, we summarize and discuss our results in Section \ref{sec:discussion}.

{\section{Observations}
\label{sec:obs}}

The UKIRT microlensing survey uses the wide-field NIR camera (WFCAM) at
the UKIRT 3.8m telescope on Mauna Kea, Hawaii. 
The 2017 fields cover the northern bulge ($b>0$), the Galactic center, and the southern bulge (see Figure \ref{fig:fields}).
Observations of the central fields are primarily done with the $K_S$-band, with a nominal cadence of 3 epochs/night. 
In addition, $H$-band observations are taken once every 3 nights.
The northern and southern bulge fields are observed once per night with the $H$-band and once every 5 nights with the $K_S$-band. 
Each epoch is composed of sixteen 5-second co-added dithered exposures (2 co-adds, 2 jitter points, and $2\times2$ microsteps).
The UKIRT dithered images are reduced, astrometrically calibrated, and stacked by the Cambridge Astronomy Survey Unit (CASU; \citealt{Irwin.2004.A}).
The light curves of all sources are then extracted using two methods -- (a) 2MASS-calibrated soft-edge aperture photometry
(standard CASU individual image catalogs; \citealt{Hodgkin.2009.A}), and (b) 2MASS-calibrated PSF photometry using SExtractor \citep{Bertin.1996.A} and PSFEx \citep{Bertin.2011.A}. The latter is better for detecting and measuring faint objects in our crowded bulge fields.

We searched the full 2017 dataset for microlensing events using the new event detection algorithm of \cite{Kim.2018.A}.
Among the events found (the full analysis will be presented in a subsequent paper), we identified UKIRT-2017-BLG-001.
The event lies inside one of our central fields, at equatorial coordinates (RA,Dec)$_{\rm J2000.0}$ = (17:46:36.98,-29:12:40.9), and Galactic coordinates $(l,b)$ = (-0.12, -0.33).

After the discovery of the event and the anomaly over its peak, we extended the observations of the field covering the event for additional two weeks beyond the main 2017 campaign.
However, these observations were only with one pointing (toward the specific field) unlike the standard observational sequence of a full tile (covering a continuous area with four pointings).
This results in a different procedure for deriving the sky frames that are used to correct for additive artifacts such as scattered light, residual reset anomaly, and illumination-dependent detector artifacts (for more details see the UKIRT/WFCAM technical website\footnote{\url{http://casu.ast.cam.ac.uk/surveys-projects/wfcam/technical/sky-subtraction}}).
The sky estimation for single pointings leaves some residual object flux and thus introduces a small systematic bias that progressively increases at faint magnitudes.
In order to incorporate this systematic effect in the PSF photometry light curve, we measured the distribution of flux offset using all of the sources near the event ($<2'$), for each epoch in the additional two weeks compared to the mean flux during the main survey. We then estimate the systematic error per epoch using the flux offset dispersion after subtracting the mean flux offset dispersion of the main survey. Finally, we add (in quadrature) this systematic error to the reported PSF flux errors.

For the modeling of the event (Section~\ref{sec:model}) we use the PSF photometry of the full dataset, while for the CMD that is used to derive the source properties (Section \ref{sec:cmd}), we use the PSF catalog from the main survey only.
Based on the RMS distribution of the PSF photometry for all of the sources in our field, we add (in quadrature) a minimum error, $e_{\min}=0.015$, to the pipeline reported errors,
in order to compensate for unrealistically small Poisson flux errors when the event is bright. We note that this is an empirical correction and is not driven by the fit for the model.

{\section{Light curve modeling}
\label{sec:model}}

The light curve (Figure~\ref{fig:lc}) has a clear and long anomaly ($\sim$8 days) just after the peak of a moderate-magnification microlensing event.
The anomaly starts with a small deviating rise above the expected single-lens model, followed by a sharp fall. Then the event rises again and continues to fall approximately 
as expected by a single-lens microlensing event. The combination of these features suggests a source crossing over a planetary resonant caustic, exiting at the ``back'' side of the caustic, opposite
to the planet (and thus the sharp fall below the single-lens model), followed by a cusp approach (see Figure~\ref{fig:caustics}).

We model the event using seven geometric parameters to calculate the magnification, $A(t)$, of a binary-lens system.
These are the three point-lens parameters $(t_0,u_0,t_\e)$ \citep{Paczynski.1986.A}, and three parameters for the companion: the mass ratio $q$,
the instantaneous scaled projected separation $s$, and the angle $\alpha$ (measured counter-clockwise) between the source trajectory
and the binary axis in the lens plane. The seventh parameter is the scaled angular source size $\rho= \theta_*/\theta_{\rm E}$.
To calculate the model magnifications near and during the anomaly we employ contour integration \citep{Gould.1997.A} with 10 annuli 
to allow for limb darkening. We adopt linear coefficients  $u_H = 0.3895$ and $u_K =0.3324$
\citep{Claret.2011.A}, based on the source type derived in
Section~\ref{sec:cmd}\footnote{We note that the best fit is insensitive to the exact limb darkening coefficient for giant stars.}.  Far from the anomaly we employ limb-darkened
multipole approximations \citep{Gould.2008.A,Pejcha.2009.A}.
Finally, each dataset has two flux parameters representing the source ($f_{s,H/K_S}$) and
any additional blend ($f_{b,H/K_S}$): $F_{H/K_S}(t) = f_{s,H/K_S} A(t) + f_{b,H/K_S}$.

To identify initial possible solutions we search over a grid of mass ratios ranging $q=0.0001-1.0$
and separations $s=0.1-1.4$, fully containing the relevant parameter space of resonant caustics.
For each grid point we initiate a set of Markov-chain Monte-Carlo (MCMC) chains with
all possible angles $\alpha$. We find an isolated single minimum centered on $(s,q)\sim(1.02,0.001)$ (see Figure~\ref{fig:grid})
and initiate from this point a full MCMC analysis (with all parameters free) to find the best-fit model.

The source of the event is faint, with $K_{S,s}\gtrsim16$, which is very close to the survey limiting magnitude (see Figure~\ref{fig:cmd}).
Moreover, our data do not completely cover the baseline of the event.
\cite{Yee.2012.A} showed that some of the standard microlensing parameters, in particular $t_E$ and $f_s$, can be poorly constrained for such events with faint sources. However, they introduced the microlensing invariants $(t_0,t_{\rm eff}\equiv u_0 t_E, t_*\equiv \rho t_E, q t_E, f_{\rm lim}\equiv f_s/u_0)$, which are well-constrained from the region near the peak and anomaly of the event.
Table~\ref{tab:model} gives the values and uncertainties of the best-fit parameters and the microlensing invariants.

We further try to include in the model microlensing parallax or orbital motion of the binary-lens system.
The inclusion of these higher-order effects does not significantly improve the fit (which requires two additional parameters for each effect) with $\Delta\chi^2=5$ when including only microlens parallax, $\Delta\chi^2=9$ when including only orbital motion, and $\Delta\chi^2=14$ when including both effects.
We do not consider either of these as a detection because they all require significant negative blending (which is unphysical) and
systematics at that level are well known in microlensing.
Nevertheless, we can set a conservative upper limit ($\Delta(\chi^2-\chi^2_{\rm best})<30$) on the microlensing parallax of $\pi_{\rm E}<0.7$.
In Section~\ref{sec:phys} we use this limit when deriving the physical properties of the system.

{\section{Source properties}
\label{sec:cmd}}

A standard way to derive the intrinsic source properties is by measuring the offset between the source position on a CMD
and the centroid of the red clump \citep{Yoo.2004.A}.
A fundamental assumption for this method is that the source and the red clump are both behind the same dust column.
The CMD is thus usually constructed using stars near ($\lesssim$2$'$) the target, as spatial extinction variations toward the bulge
are usually on larger scales. However, in the case of our event there are several challenges with this method.

The first challenge is the observed source properties. The entire range of $(H - K_S)$ colors for giants (which our source likely is, as we show below) is narrow, with only 0.24 mag difference between G0 to M7 giants. Even for dwarfs the range is only 0.38 mag between B8 and M6 dwarfs \citep{Bessell.1988.A}\footnote{We convert \citealt{Bessell.1988.A} $(H - K)_{\rm BB}$ colors to 2MASS-calibrated $(H - K_S)_{\rm 2MASS}$ colors using the relations from \cite{Carpenter.2001.A}.}.
This is because $(H - K_S)$ basically probes the Rayleigh--Jeans tail for most stars.
Therefore, a precise measurement of the source color is essential in order to derive the source properties. We determine the model-independent source $(H - K_S)_s$ color from
regression of $H$ versus $K_S$ flux as the source magnification changes \citep{Gould.2010.A}, and find $(H - K_S)_{s}=1.87\pm0.01$. (This color is in excellent agreement with the color derived from the model of $(H - K_S)_{s}=1.878\pm0.004$).
Each $H$ epoch was taken in between two $K_S$ observations on the same night, minimizing systematics and allowing us to achieve the required precision.
Next, the source is faint and, as discussed in Section~\ref{sec:model}, we do not have coverage of the event's baseline.
Therefore, the uncertainty on the source magnitude is relatively large. The $K_S$ source magnitude as inferred from the microlensing model is 
$K_{S,s}=16.07_{-0.11}^{+0.09}$. The model also indicates a fainter blend flux, with 5$\sigma$ upper limit of $K_{S,b}>17.7$. 
We use this measurement later as an upper limit on the lens flux when we estimate the physical properties of the planetary system (see Section~\ref{sec:phys}).
For completeness, we note that no centroid shift as a function of magnification was detected.

The second and more prominent challenge is the determination of the extinction and reddening using the red clump centroid.
There are clear dust clouds in the region around our event (see Figure~\ref{fig:fchart}), which cause large spatial reddening variations on scales smaller than the standard $\lesssim$2$'$ region. We avoid obvious dust stripes and use stars in a 1.6 arcmin$^2$ box around the event, as marked in Figure~\ref{fig:fchart}, to construct the CMD (Figure~\ref{fig:cmd}). We measure the red clump centroid following the procedure described in \cite{Nataf.2013.A} and find $(H - K_S,K_S)_{\rm cl}=(1.63,14.63)$.
The number of red clump stars used, as derived from the fit, is $N_{RC}=197\pm14$, which is sufficient for a reliable measurement of the red clump.
By comparing the apparent clump centroid to the
intrinsic centroid of $(H - K_S,K_S)_{\rm cl,0}=(0.15,12.95)$ toward the event's location (assuming a distance of 8.17 kpc;  \citealt{Nataf.2013.A,Nataf.2016.A,Hawkins.2017.A}) we find that 
the mean extinction and reddening toward the event are $\langle A_{K_S}\rangle=1.68$ and $\langle E_{H - K_S}\rangle=1.48$, respectively.
We note that the reddening-to-extinction ratio that we find is non-standard, and we investigate this further in Section~\ref{sec:ext_law}.

However, both the color and magnitude dispersions of the red clump stars are large, suggesting large extinction and reddening dispersions (and thus large uncertainties on the intrinsic source properties) as well as a possibly wide distance dispersion of clump stars. The observed clump color dispersion is $\sigma_{(H - K_S)_{cl}} = 0.16$ (as derived from the clump centroid fit; see \citealt{Nataf.2013.A} for details), which combines the intrinsic clump color dispersion and reddening dispersion. 
In order to estimate the intrinsic dispersion we apply the red clump centroid procedure to several of our low-Galactic-latitude UKIRT fields that have low and uniform extinction. From the clump color dispersion on these fields we find that the intrinsic clump color dispersion is only 0.04
and thus the uncertainty on the reddening (e.g., the reddening dispersion) in the event's field is $\sigma_{E_{H - K_S}} = 0.15$. This large uncertainty, if taken as is, implies that the intrinsic color of the source cannot be well constrained, as the 
full range of colors for giant stars is within $2\sigma_{E_{H - K_S}}$ and the full dwarf color range is within $2.5\sigma_{E_{H - K_S}}$.
The observed clump magnitude dispersion, $\sigma_{K_{S,cl}} = 0.36$, is even higher.
Intrinsic dispersion, extinction dispersion, and distance modulus dispersion all contribute to the total magnitude dispersion (the dispersion due to photometric noise is negligible).
The intrinsic magnitude dispersion is 0.17 \citep{Hawkins.2017.A}, and the extinction dispersion can be estimated from the reddening dispersion and the mean extinction-to-reddening ratio to be $\sigma_{A_{K_S}} = 0.17$. Thus, the distance modulus dispersion to the clump is 0.28. While this value is larger than typical toward Baade's window \citep{Nataf.2013.A}, the event is located very close to the Galactic center (and thus the Galactic plane), implying a large geometrical dispersion due to 
the fact that there are significant contributions from both bulge and far disk stars (see more details in Section~\ref{sec:far_disk} below).

The large differential reddening implies that the basic assumption that the source is behind the same dust as the clump is not reliable, as the extinction itself is not uniform across the field. Moreover, assuming that the reddening and extinction to the source are the mean reddening and extinction in the field, $\langle E_{H - K_S}\rangle=1.48$ and $\langle A_{K_S}\rangle=1.68$, yield intrinsic source properties of $[(H - K_S),K_S)]_{s,``0''} = (0.39,14.39)$. While the magnitude suggests a giant source, the color is 
0.05 redder than an M7 giant (which is already very rare, as suggested by the small intrinsic dispersion of clump star colors). This implies that the source is suffering from higher reddening than the mean clump.
While we cannot use the usual method to derive the source properties with small uncertainties, we can use it to set possible limits on the source properties.
We assume that the extinction and reddening toward the source are $\langle A_{K_S}\rangle+\Delta A_{K_S}$ and $\langle E_{H - K_S}\rangle+\Delta E_{H - K_S}$, respectively.
In order to estimate the boundaries of the possible source angular size range, we take two extreme cases of giant sources.
As the reddest limit, we assume an M7 giant with $(H - K_S)=0.34$, thus suffering from additional reddening $\Delta E_{H - K_S}=0.05$ and correspondingly additional extinction $\Delta A_{K_S}=0.06$.
As the bluest limit, we assume a G0 giant with $(H - K_S)=0.10$, thus suffering from $\Delta E_{H - K_S}=0.29$ and $\Delta A_{K_S}=0.33$.
For each case, we derive the source intrinsic magnitude by accounting for the total extinction. We then use the $(V-K)$ colors \citep{Bessell.1988.A} corresponding to the source spectral type and the surface brightness to angular source size relation of \cite{Kervella.2004.A} to infer that the source angular size is in the range $3.3 <\theta_*/\muas < 8.5$.

{\subsection{Far disk source?}
\label{sec:far_disk}}

The large differential reddening and the ``too red'' source color imply a non-uniform extinction toward each individual star, and in particular the source.
These can be due to either spatial differential reddening on the scale of a few arcseconds (even at the apparently uniform box selected), a different accumulated dust column along the line of sight toward sources at different distances, or a combination of both. We first investigate the possibility of variations of dust columns due to source distances, which we find suitable to fully explain both problems. Nevertheless, in Section \ref{sec:spatial} we try to examine the possible spatial differential reddening.

The event is located at Galactic latitude $b=-0.33^\circ$, corresponding to $\sim50$ pc below the Galactic plane at the distance of the Galactic bulge ($\sim8$ kpc), or
115 pc at the extreme far disk (20 kpc). Therefore, the Galactic thin disk population can have a significant contribution to the number of observed RC stars on the CMD, because the scale height of the Galactic thin disk is 300 pc \citep{Juric.2008.A}. This can partially explain the observed scatter.
In addition, the dust scale height is 120 pc \citep{Jones.2011.A}. First, this can explain the large differential reddening within the bulge. Second, stars in the far disk will be significantly more extinguished. The \cite{Marshall.2006.A} 3D Galactic interstellar extinction model, which combines 2MASS data \citep{Cutri.2003.A} and the Besan{\c c}on Galactic model \citep{Robin.2003.A}, suggests an additional extinction toward the event of $\Delta A_{K_S}\approx0.4$ behind the bulge (between 8.5 and 12 kpc where they are limited by the data). The \cite{Green.2015.A} 3D Galactic dust map, which combines the 2MASS and Pan-STARRS1 \citep{Kaiser.2010.A} datasets, suggests a $\sim$$15\%$ increase in reddening (corresponding to $\Delta E_{H - K_S}\approx0.25$) between 8 and 16 kpc.
However, both of these maps \citep{Marshall.2006.A,Green.2015.A} run out of stars before or around the distance to the Galactic center, and thus they probably underestimate the true amount of additional extinction on the far side of the Galaxy. 
In conclusion, the combined effects of the significant thin disk population and the accumulated dust result in the large scatter of the apparent magnitude of the RC stars in the field.

The source distance and position on the CMD can similarly be well explained by the above arguments.
First, we estimate the general probability of a far-disk source with $15<K_{S,s}<17$ toward the event’s coordinates
using a new Galactic population synthesis model (Huston et al., in preparation)\footnote{The model assumes a bulge with an E3 density distribution from \cite{Dwek.1995.A}, with parameters estimated by \cite{Cao.2013.A} using OGLE-III clump giants. The bar has an angle of 29$^\circ$ to the line of sight, and kinematics that match the BRAVA survey \citep{Howard.2009.A}. The disk density and age distribution follows that of \cite{Robin.2003.A}. Stellar properties are derived from MIST version 1.1 solar metallicity isochrones \citep{Dotter.2016.A,Choi.2016.A}.}
and Monte-Carlo integration of the event rate \citep{Penny.2013.A,Awiphan.2016.A}.
We find that the contribution of far-disk sources, at distances $D_s>11$ kpc, is 50\% of the overall general source probability.
Next, as we found above, the source is $\Delta (H - K_S)=0.25$ redder than the bulge mean clump. 
If, for example, the source is a {\it typical red clump star at the far disk}, then this difference is due to additional reddening $\Delta E_{H - K_S}=0.25$. Using the mean extinction-to-reddening ratio derived by the mean clump position, this corresponds to $\Delta A_{K_S}\approx0.28$. The remaining difference between the source magnitude and the mean clump from the change in distance modulus is thus 1.16. This implies a source that is 1.71 times more distant than the mean clump, corresponding to $D_s\approx14$ kpc.
As we showed above, Galactic dust models and observations suggest that, at such distance, the source is likely to suffer from the observed additional extinction and reddening.
The distance could be somehow less than 14 kpc, if the source is intrinsically fainter than the clump or
if the dust on the other side of the Galaxy has an extinction curve that is closer to the standard law than the dust between the Sun and the red clump.
Later, in our Bayesian analysis (Section~\ref{sec:phys}) we find that the source is most likely in the far disk ($D_S = 11.2^{+3.6}_{-2.6}$ kpc), and thus we conclude that this is probably the correct explanation.

{\subsection{Spatial differential reddening}
\label{sec:spatial}}

The field around the event is severely impacted by dust stripes. While we constructed our CMD by selecting a box without any obvious dust clouds, it is still possible that the large differential reddening is due to local spatial variations on the scale of a few arcminutes. We tried to examine this possibility by extracting mutli-color photometry of the stars around the event. By combining extinction ratios (either by using the red clump stars for each band or through color-color relations) and the colors of stars, we can estimate the extinction toward individual stars. This will allow, in principle, the derivation of a spatial reddening map. 

We obtained $grizY$ bands catalog of stars near the event using data from the DECam Plane Survey \citep{Schlafly.2018.A}, with corresponding AB limiting magnitudes of 23.9, 23.0, 22.6, 21.9, and 21.2. In addition, we extracted a $Spitzer$ 3.6$\mu m$ catalog of stars near the event from an image taken as part of the GLIMPSE Proper project \citep{Benjamin.2015.A} using the new IRAC crowded-field photometry algorithm \citep{Calchi.2015.B}, and calibrating it to \cite{Ramirez.2008.A}. The limiting magnitude is 15.4.
Finally, we extracted VISTA Variables in the V{\'i}a L{\'a}ctea (VVV; \citealt{Minniti.2010.A}) $J$-band catalog\footnote{We note that our UKIRT data are deeper than the $H$ and $K_S$ VVV catalogs.}, with a limiting magnitude of 19.4.
Unfortunately, most of the stars in our field are not detected in the DECam $grizY$ and VVV $J$ catalogs (see Figure~\ref{fig:ext_law}). In particular, the bulge mean red clump is not fully covered in all of these bands. Therefore, we cannot derive the spatial reddening map using the method suggested above, and thus further observations are required in order the explore this viable explanation, as we discuss in Section~\ref{sec:discussion}.
Nevertheless, we use the limiting magnitudes of these catalogs to study the extinction curve toward this field in Section \ref{sec:ext_law}.

{\section{Planetary system physical properties}
\label{sec:phys}}

The angular Einstein radius and the relative geocentric proper motion between the source and the lens can be derived from the light curve model and CMD analysis,
$\theta_{\rm E}=\theta_*/\rho$ and $\mu_{\rm geo}=\theta_{\rm  E}/t_{\rm E}$. Applying the limits derived above on $\theta_*$ then gives
\begin{equation}
0.44 <\theta_{\rm E} [{\rm mas}] < 1.39
\qquad\qquad
1.5 <\mu_{\rm geo} [{\rm mas/yr}] < 5.5.
\end{equation}

Unfortunately, even using these limits we cannot determine directly the physical properties of the planetary system,
as the microlens parallax was not detected. We therefore estimate them using a Bayesian analysis
that incorporates the limits on $\theta_{\rm E}$ and $\mu_{\rm geo}$
into a Galactic model.
We follow the procedures described in \cite{Shvartzvald.2014.A} and adopt the Galactic model of \cite{Han.1995.A,Han.2003.A}, which reproduces well the observed statistical distribution of properties of 
microlensing events.
For the source probability, we apply conservative color and magnitude limits of $15<K_{S,s}<17$ and $(H - K_S)_s>1.3$.
We furthermore set two limits on the mass-distance relation of the lensing system. The first is from the microlens parallax
\begin{equation}
 \pi_{\rm E}\equiv \sqrt{\kappa M \pi_\rel}<0.7
 \qquad
\kappa\equiv {4 G\over c^2\au}\simeq 8.14{{\rm mas}\over M_\odot}.
\end{equation}
Here $\pi_\rel = \au(D_L^{-1}-D_S^{-1})$ is the lens-source relative parallax.
The second limit is an upper limit on the lens flux by using the 5$\sigma$ upper limit on the blend flux that was derived from the model, $K_{S,l}\geq K_{S,b}>17.7$.
We use 5 Gyr Padova isochrones \citep{Bressan.2012.A,Marigo.2017.A} and assume (as a conservative limit) that the lens is behind the overall dust column toward the bulge
to convert the flux limit to a mass-distance limit.
This limit alone already gives an upper limit on the planet mass (if all the blend flux is attributed to the lens), because at this age the maximal host mass is $\lesssim1.25M_\odot$,
and thus the planet mass will be  $<2.7 M_J$.
We note that if the host is a bulge star, with age of $\approx10$ Gyr, the limits on the host and planet masses are $\lesssim1.05M_\odot$
and $<2.2 M_J$, respectively. However, as the event is within one scale height of the thin disk throughout the Galaxy, we use the 5 Gyr isochrones to derive the limit on the lens flux for all distances, as there is a non-negligible probability that the lens might be a part of the disk population even at 6--10 kpc.

The 2D posterior distribution of the host mass and distance is shown in Figure~\ref{fig:bayeisan}, as well as the limits on $\theta_{\rm E}$, $\pi_{\rm E}$, and the lens flux.
We infer that the host is a $0.81^{+0.21}_{-0.27} M_\odot$ dwarf, likely in the Galactic bulge at $6.3^{+1.6}_{-2.1}$ kpc.
Using the mass ratio and scaled instantaneous projected separation from the model, we find that the planet mass is $1.28^{+0.37}_{-0.44} M_{J}$, and it orbits its host beyond the snowline at a projected separation of $4.18^{+0.96}_{-0.88}$ AU.
The estimated physical properties are summarized in Table~\ref{tab:phys}.

One of the main uncertainties in our Galactic model is the source distance. However, we note the the posterior probability for the lens mass (and thus the planet mass) is weakly dependent on the source distance. For a source at 8 kpc the mass is only 10\% lower than for a source at 15 kpc, well within the range of our uncertainty.
The lens distance, and consequently the projected separation between the planet and its host, is more sensitive to the source distance, with a difference of 45\% between a bulge source and a far disk source. Future observations can resolve this and give a better estimation of the source distance, as we discuss below in Section~\ref{sec:discussion}.

{\section{Extinction law}
\label{sec:ext_law}}

Recent studies of dust properties in the inner bulge suggest deviations from the standard extinction law \citep{Nataf.2016.A,AlonsoGarcia.2017.A}.
The extinction coefficient measured in our field, $\langle A_{K_S}\rangle/\langle E_{H - Ks}\rangle = 1.14$, is lower than the value of 1.37 from \cite{Nishiyama.2009.A}. This offset is a disconcertingly large $\sim$0.4 mag in $K_{S}$. The offset relative to the value of  $A_{Ks}/E_{H - Ks} \sim2.07$ predicted by \cite{Fitzpatrick.1999.A} is even larger -- about 1.4 mag in $K_{s}$. 

The magnitude of the red clump can also be predicted in the $YJ$ bands by assuming the color-color relations from \cite{Nataf.2016.A} and the extinction coefficients from \cite{Fitzpatrick.1999.A}. These are $Y_{RC} \approx 14.63+1.63+(7.35-3.12) \approx 20.49$ and $J_{RC} \approx 14.63+1.63+(5.44-3.12) \approx 18.58$, where we
add the corresponding intrinsic color offsets and reddening offsets,
respectively, to the observed clump $H$-band magnitude.

Given that the limiting magnitudes for the DECam and VVV datasets are, respectively, $Y \sim 21.2$ and $J \sim 19.4$ (see Section~\ref{sec:spatial} above), the red clump should be clearly detected in each of these bandpasses by a wide margin of 0.7--0.8 magnitudes, respectively. However, the photometry in both cases barely covers the red clump.
Figure \ref{fig:ext_law} shows the $HK_{s}$ CMD of our field (gray points), with the sources detected in $Y$ shown as yellow circles in the left panel, and the points detected in $J$ shown as purple in the right panel.
We conclude that the extinction toward this field is steeper than standard (as defined by \citealt{Fitzpatrick.1999.A}). 

{\section{Discussion}
\label{sec:discussion}}

We have presented the discovery of a roughly Jupiter/Sun mass ratio planet.
The system likely resembles a version of the Jupiter/Sun system, with the host being a G/K dwarf slightly less massive than the Sun, located in the Galactic bulge.
The event was detected as part of our UKIRT microlensing survey, which has the primary goal of deriving the NIR event rate toward the {\it WFIRST} target field region.
As such, the survey is designed to have a $\sim$daily cadence, which in principle is not optimal for the detection of short planetary anomalies.
However, the perturbation was due to a resonant planetary caustic, and the event had a long timescale, so the cadence was sufficient to detect the planetary anomaly.
This event contributes to the set of interesting planetary microlensing events that were discovered through resonant caustic perturbations.
The resonant caustic parameter space is relatively wide for giant planets. For $q=10^{-3}$, resonant caustics exist for instantaneous projected separations of $0.93<s<1.15$, corresponding to 2.6--3.3 AU for a typical Einstein radius.
However, 
the range in projected separations for which resonant caustics exist scales as $q^{1/3}$, and thus resonant caustics are less important for lower mass ratios $q$.

The field around the event suffers from high and differential extinction that creates a challenge for deriving the intrinsic source properties, and thus the physical properties of the planetary system. Follow-up observations of UKIRT-2017-BLG-001Lb can reduce these uncertainties in several ways.
First, deep high-angular resolution imaging of the field in the infrared would enable an accurate measurement of the extinction toward each star (including the source) using the Rayleigh-Jeans color excess method \citep{Majewski.2011.A}. This can be done from the ground with Keck in the NIR or, preferably, with {\it JWST} in mid- and near-infrared bands allowing for a wider spectral range. Second, {\it JWST} could also measure a  medium-resolution spectrum of the source star in the NIR to estimate its temperature and improve the estimate of the angular Einstein radius (the upgraded NIRSPEC on Keck will potentially have the sensitivity for a low-resolution spectrum of the source in $H$-band).
Third, the high-resolution image will resolve out possible unrelated sub-arcsecond blend stars around the target (see, e.g., \citealt{Beaulieu.2018.A}), allowing the measurement of any excess flux from the target above the measured source flux. However, the blend flux from the microlensing model ($K_{S,b}>17.7$) already sets an upper limit on the excess flux.
For such faint excess the probability for a significant contribution from stars that are not the lens (i.e., companion to lens, companion to source, or ambient star)
is high (see, e.g., \citealt{Koshimoto.2017.A}).
Finally, a second high-resolution epoch of UKIRT-2017-BLG-001Lb could resolve the lens and the source and directly measure the relative lens-source proper motion and the lens flux (e.g., \citealt{Batista.2015.A}). Given that the relative proper motion is in the range $1.5 <\mu [{\rm mas yr^{-1}}] < 5.5$ (with Bayesian estimate of $\mu_{\rm geo}\sim2.5 $ mas yr$^{-1}$) the source and the lens will be sufficiently separated in 10--50 years. However, centroid shifts due to the relative lens-source proper motion (e.g., \citealt{Bennett.2015.A,Bhattacharya.2017.A}) could be detected earlier.

Future NIR microlensing surveys, such as the planned {\it WFIRST} microlensing program
or the one proposed with {\it Euclid} \citep{Penny.2013.A},
may consider avoiding such regions with large spatial extinction variations.
On the other hand, our current results \citep{Shvartzvald.2017.A} suggest that the NIR event rate is highest close to the Galactic center, and thus the field selection should balance between the high differential extinction and the high event rate.
We note that the currently planned {\it WFIRST} fields (Penny et al. in preparation) lie at lower Galactic latitudes ($b \approx -2.05$ to $-0.45$), in regions of lower extinction and lower differential extinction than UKIRT-2017-BLG-001Lb\footnote{Based on the extinctions of red clump stars measured by \cite{Nidever.2012.A},  the current {\it WFIRST} fields have quartile $K_s$ extinctions of 0.38 and 0.64.}.

The multi-band analysis of our field suggests a non-standard (steeper) extinction law. This 
supports previous suggestions of interstellar extinction law variations toward the inner bulge \citep{Nataf.2016.A,AlonsoGarcia.2017.A}.
Our ongoing UKIRT survey will enable the creation of extinction and reddening maps of the Galactic center and bulge. These, combined with deep optical surveys, will allow us to confirm and fully constrain these variations.
These results can also have important implications for observational cosmology,
as they suggest that, if non-standard extinction laws occur in external galaxies, they may lead to
systematic errors in the distances derived from studies of SNe Ia and Cepheids,
if those studies adopt standard extinction laws.

\acknowledgments
We would like to dedicate this paper to the memory of Neil Gehrels,
whose support made the 2017 UKIRT survey possible.
We thank M. Irwin for useful discussions,
D. Imel for assistance with developing cloud processing for the light curves,
and the anonymous referee for important comments that helped to improve the manuscript.
Work by YS was supported by an
appointment to the NASA Postdoctoral Program at the Jet
Propulsion Laboratory, administered by Universities Space Research Association
through a contract with NASA.
DMN was supported by the Allan C. and Dorothy H. Davis Fellowship.
This work was partially supported by NASA grants NNX17AD73G and NNG16PJ32C.
UKIRT is owned by the University of Hawaii (UH) and operated by the UH Institute for Astronomy;
when some of the data reported here were acquired, UKIRT was supported by NASA
and operated under an agreement among the University of Hawaii, the University of Arizona,
and Lockheed Martin Advanced Technology Center;
operations were enabled through the cooperation of the East Asian Observatory.
Based on data products from observations made with ESO Telescopes at the La Silla Paranal Observatory
under programme ID 179.B-2002.

\begin{table}
\centering
\caption{ Microlensing model \label{tab:model}}
\begin{tabular}{|l|c|}
\tableline
Parameter	& \\
\tableline
$t_0$ [HJD$'$]& $7916.243\pm0.022$\\[5pt]
$u_0$& $0.0303_{-0.0026}^{+0.0036}$\\[5pt]
$t_{\rm E}$ [d]& $101.0_{-9.6}^{+8.2}$\\[5pt]
$\rho$ [$10^{-3}$]& $6.64_{-0.56}^{+0.75}$\\[5pt]
$\alpha$ [rad]& $3.7036_{-0.0076}^{+0.0084}$\\[5pt]
$s$& $1.0318_{-0.0045}^{+0.0032}$\\[5pt]
$q$ [$10^{-3}$]& $1.50_{-0.14}^{+0.17}$\\[5pt]
$K_{S,s}$& $16.07_{-0.11}^{+0.09}$\\[5pt]
$K_{S,b}$& $>17.7^*$\\[5pt]
\tableline
$t_{\rm eff}$ [d] & $3.062\pm0.050$\\[5pt]
$t_*$ [d] &$0.671\pm0.016$\\[5pt]
$q t_{\rm E}$ [d] &  $0.1517\pm0.0041$\\[5pt]
$f_{lim}$ & $198.2\pm2.0$\\[5pt]
\tableline\tableline
\end{tabular}
\newline
\raggedright{HJD$'$=HJD - 2450000}
\newline
\raggedright{$^*$ 5$\sigma$ limit}
\end{table}

\begin{table}
\caption{Physical properties\label{tab:phys}}
\begin{tabular}{|l|c|}
\tableline\tableline
$M_1$ [$M_\odot$]& $0.81^{+0.21}_{-0.27}$ \\[5pt]
$m_2$ [$M_J$]&  $1.28^{+0.37}_{-0.44}$\\[5pt]
$r_\perp$\,\, [$\au$]&  $4.18^{+0.96}_{-0.88}$\\[5pt]
$D_L$ [kpc]&  $6.3^{+1.6}_{-2.1}$\\[5pt]
$D_S$ [kpc]&  $11.2^{+3.6}_{-2.6}$\\[5pt]
$\mu_{\rm geo}$ [mas yr$^{-1}$]&  $2.34^{+0.84}_{-0.42}$\\[5pt]
\tableline\tableline
\end{tabular}
\end{table}

\begin{figure}
\plotone{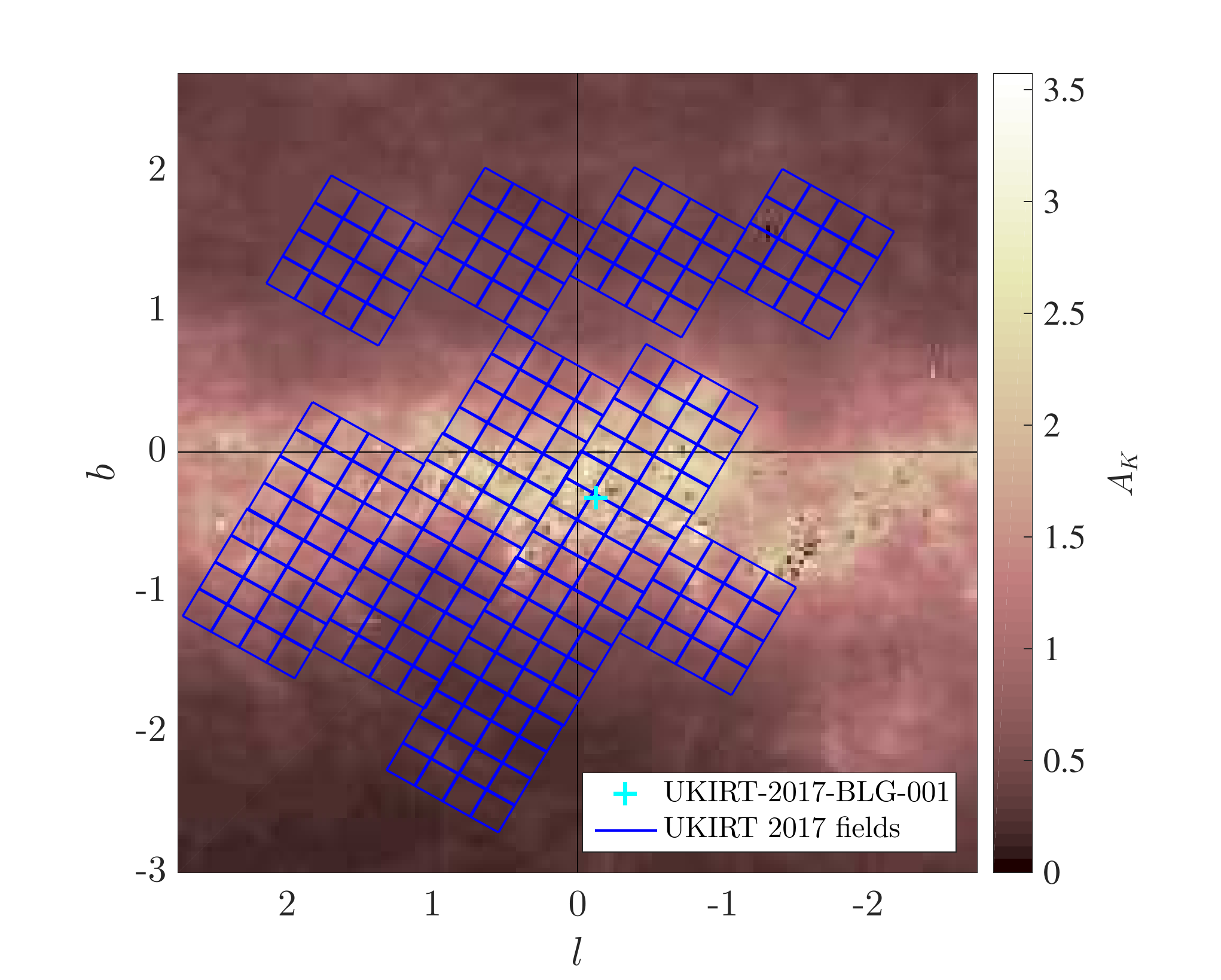}
\caption{UKIRT 2017 microlensing survey fields (blue), plotted over the $A_K$ extinction map from \cite{Gonzalez.2012.A}.
The cyan plus marks the location of the event UKIRT-2017-BLG-001.
}
\label{fig:fields}
\end{figure}

\begin{figure}
\plotone{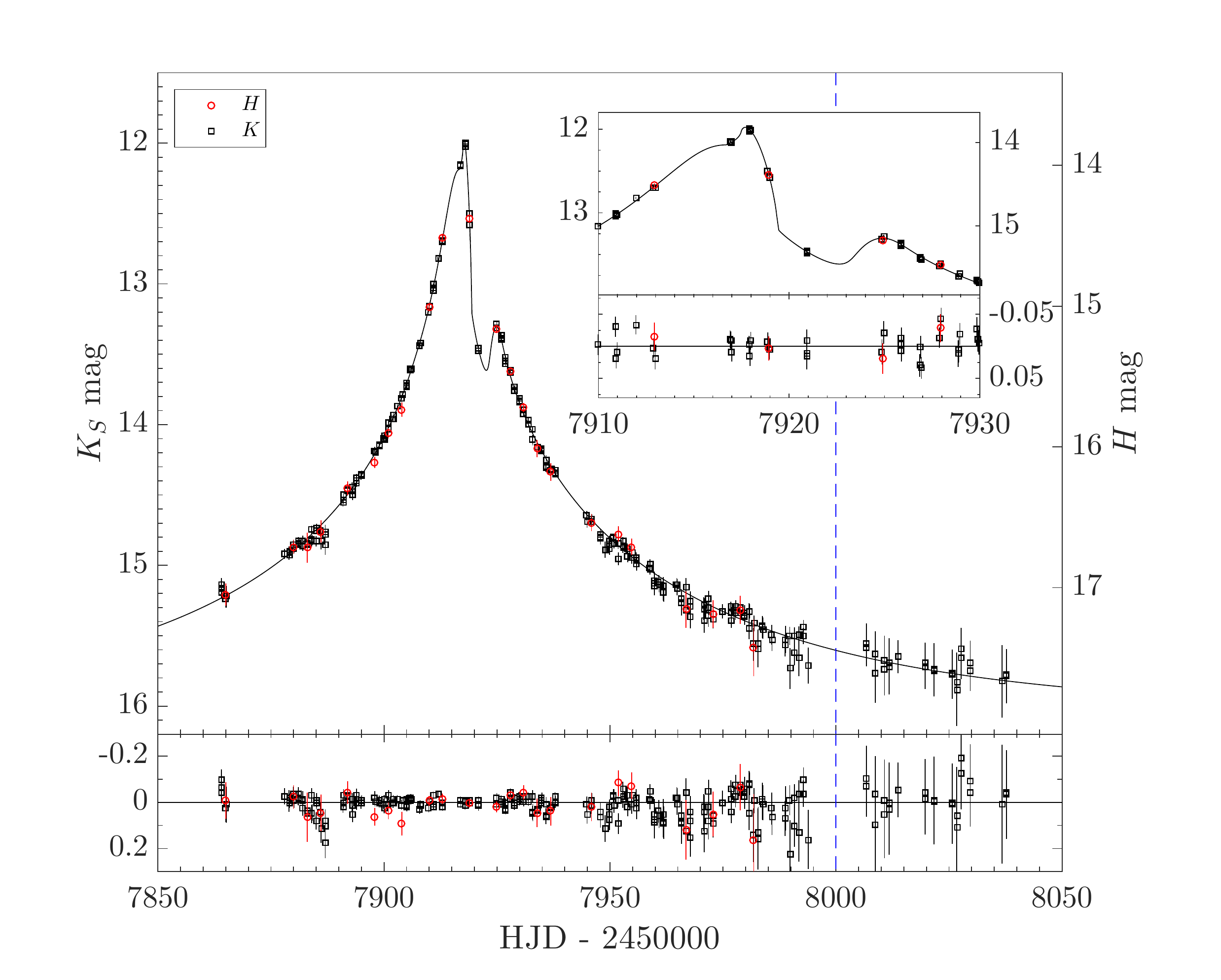}
\caption{Light curve of UKIRT-2017-BLG-001 in $K_S$ (black squares and left horizontal axis) and $H$ (red circles and right horizontal axis).
The best-fit planetary model is shown in black.
The anomaly (inset) over the peak is clear and covered sufficiently well with our 3 epochs/night cadence.
The vertical blue dashed line indicates the end of the main survey, beyond which we include systematic errors due to poor sky estimations.
}
\label{fig:lc}
\end{figure}

\begin{figure}
\centering
\includegraphics[width=0.5\textwidth]{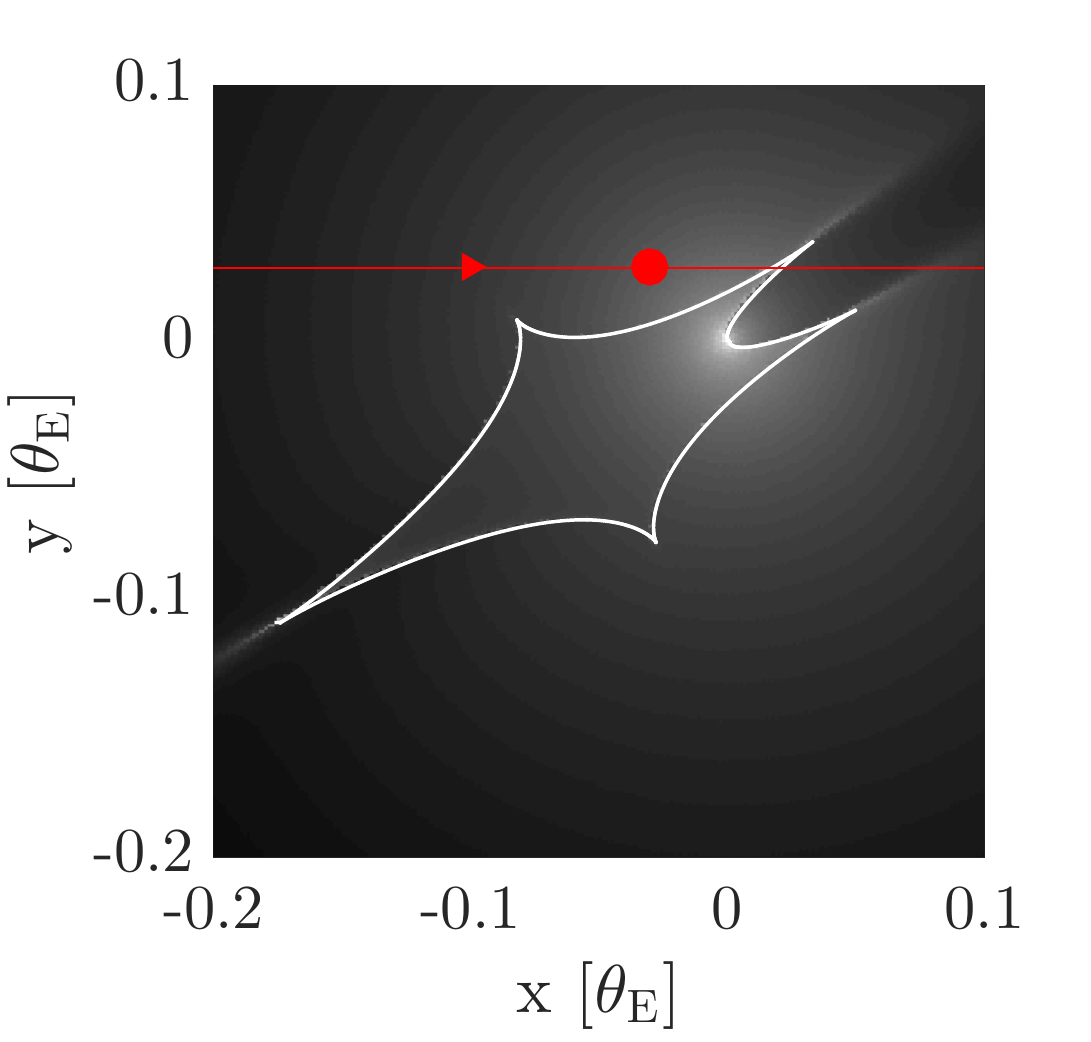}
\caption{Magnification map of the event. The source trajectory (red line) over the resonant caustic (white curve). The red circle indicates the source size.
The length of the caustic (long axis) is larger than 0.2$\theta_\e$, showing the large cross section of giant planet resonant caustics.
The width is $\sim$0.1$\theta_\e$ and the event timescale is $t_\e\approx100$ days, suggesting a typical anomaly duration of $\sim$10 days.
}
\label{fig:caustics}
\end{figure}

\begin{figure}
\plotone{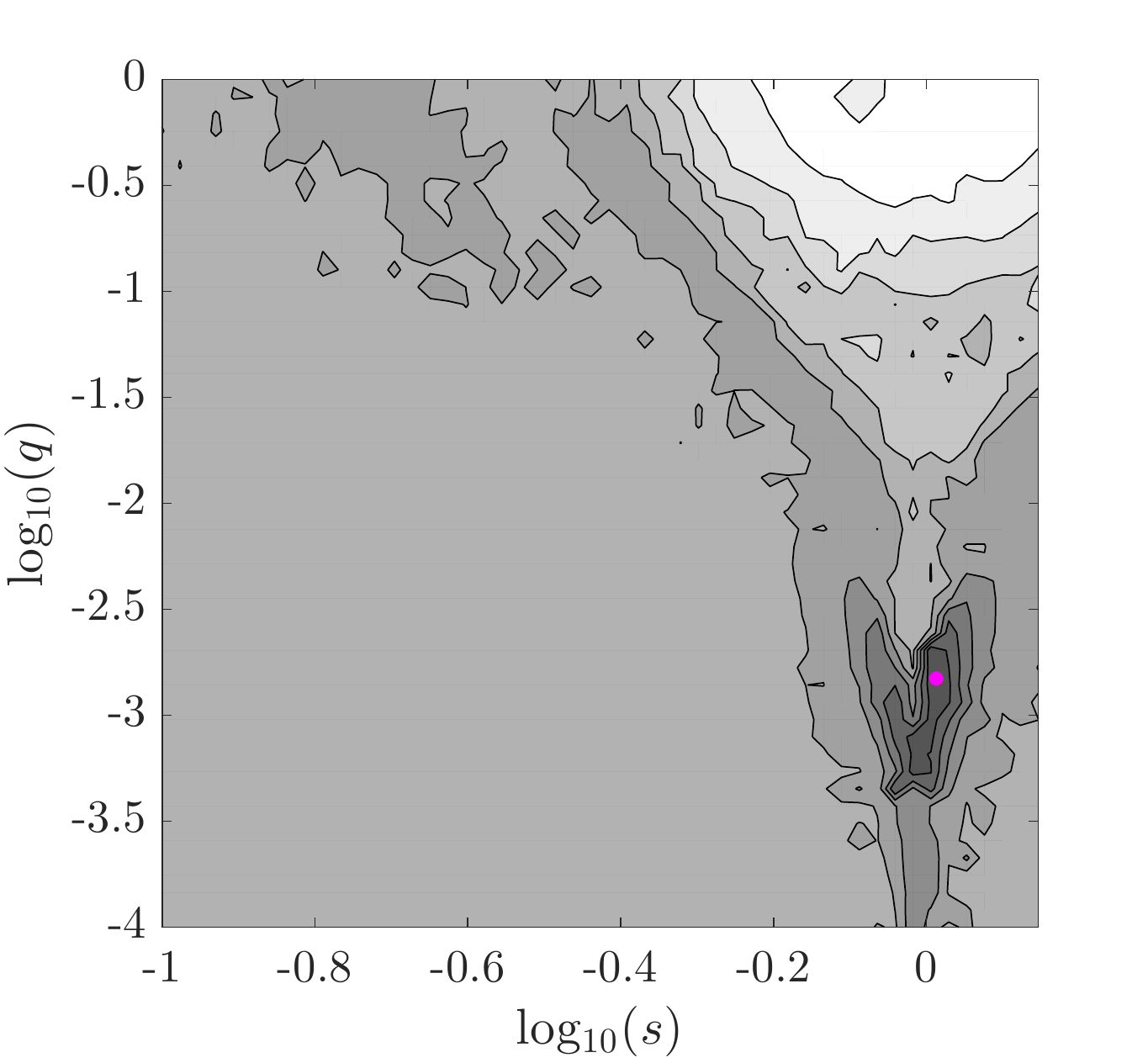}
\caption{Results from the grid search on the $s$-$q$ (scaled separation, mass ratio) plane. The values of the best-fit solution are marked as a magenta circle.
A single isolated minimum is clearly detected.
The contours indicate a steep surface, where all results with $\Delta\chi^2<100$ compared to the best fit are within the inner contour and results with $\Delta\chi^2<500$ and $\Delta\chi^2<1000$ are within the second and third contours, respectively.
}
\label{fig:grid}
\end{figure}

\begin{figure}
\plotone{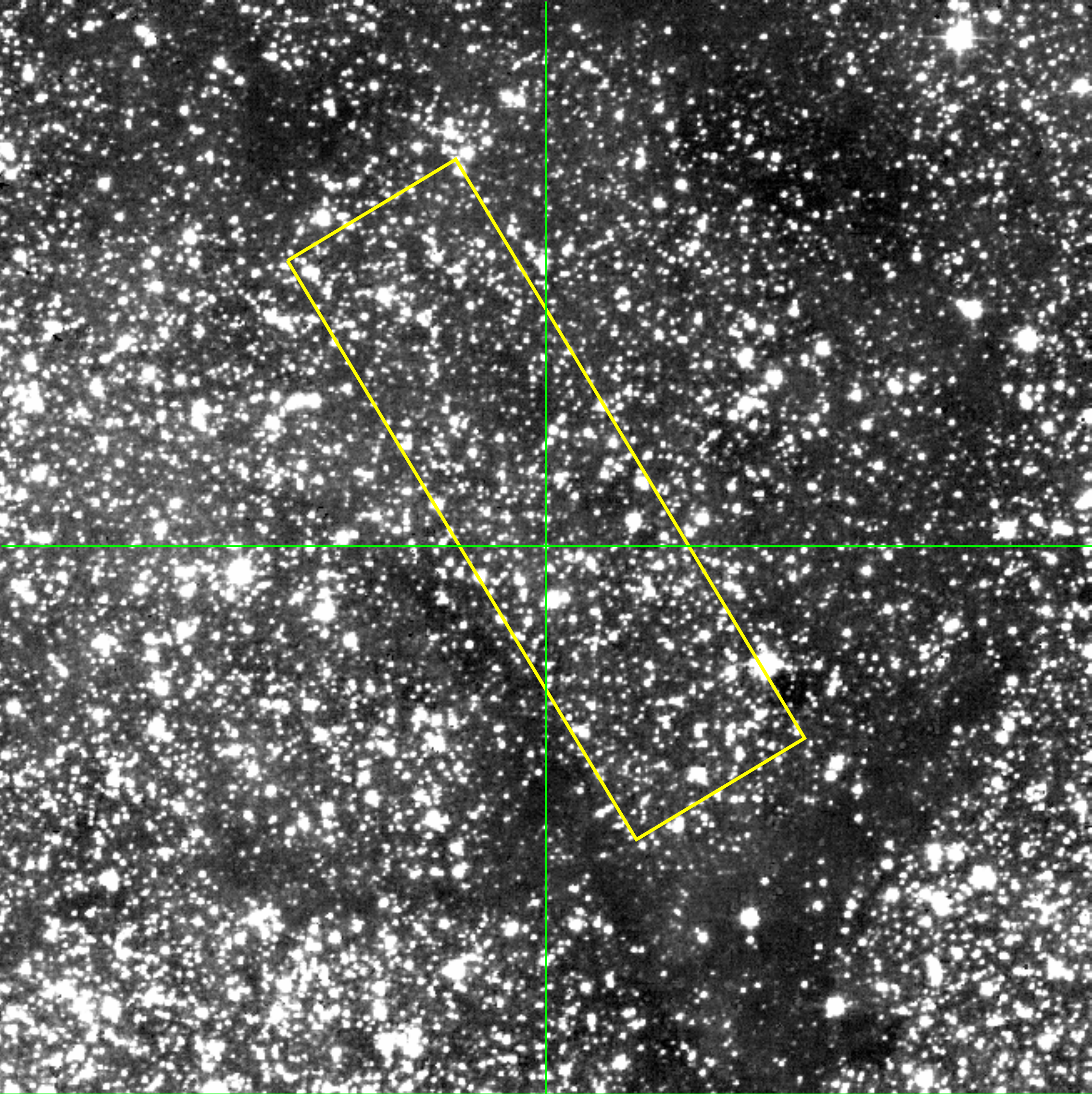}
\caption{UKIRT 3.8' x 3.8' $H$-band image of the field around the event (marked with a cross-hair).
Dust ``stripes'' are clearly seen around the target, indicating the high differential reddening in the field.
The yellow box marks the region used to construct the CMD. 
}
\label{fig:fchart}
\end{figure}

\begin{figure}
\plotone{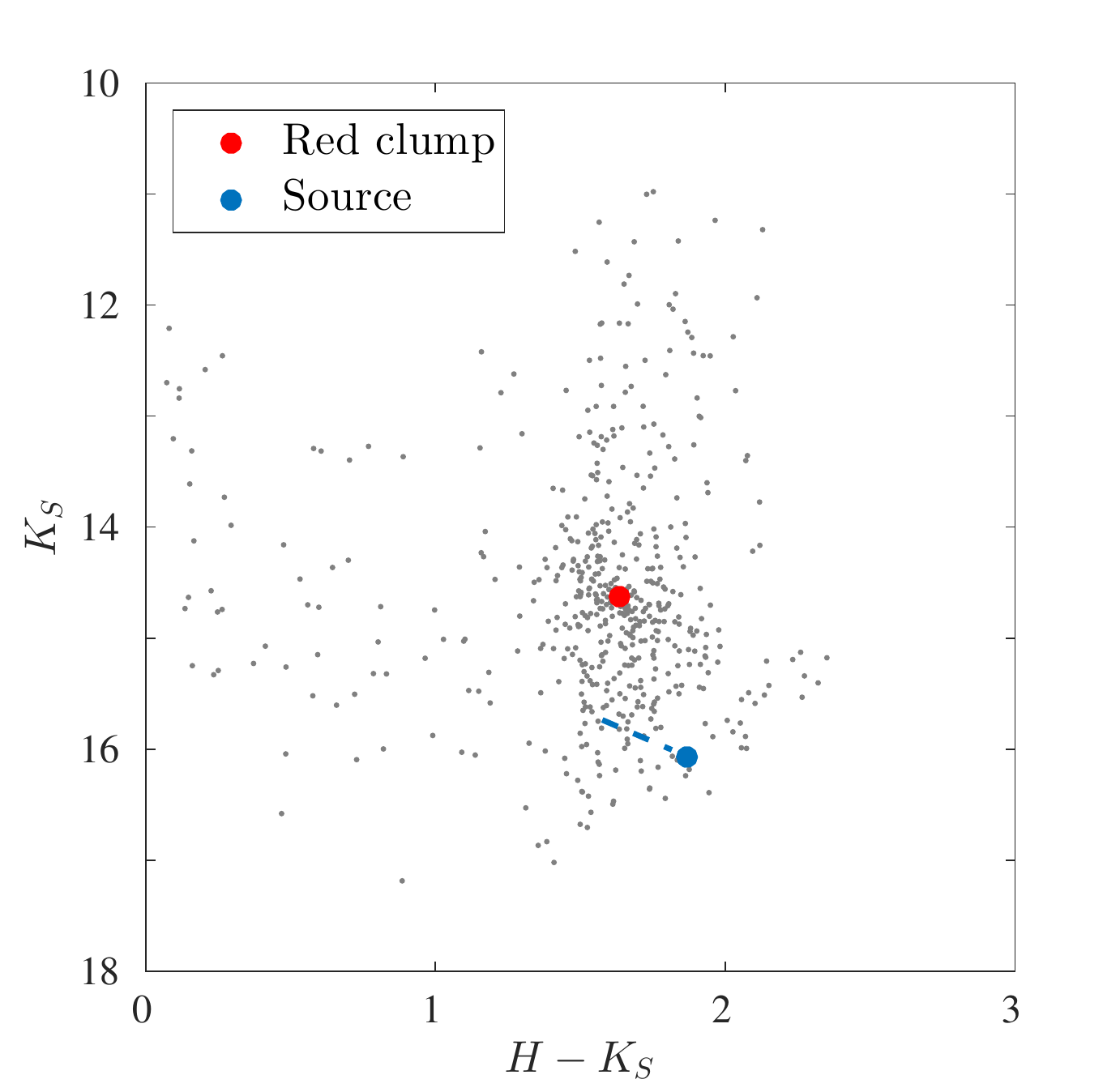}
\caption{Color-magnitude diagram of stars around the event. The red circle indicates the centroid of the red clump,
with the relatively large dispersion around it indicating the high differential reddening.
The large color offset between the source (blue) and the clump suggests that the source suffers from additional reddening.
The blue dashed line indicates the source position as if it were behind the same dust column as the clump for the range of possible giant colors.
}
\label{fig:cmd}
\end{figure}

\begin{figure}
\plotone{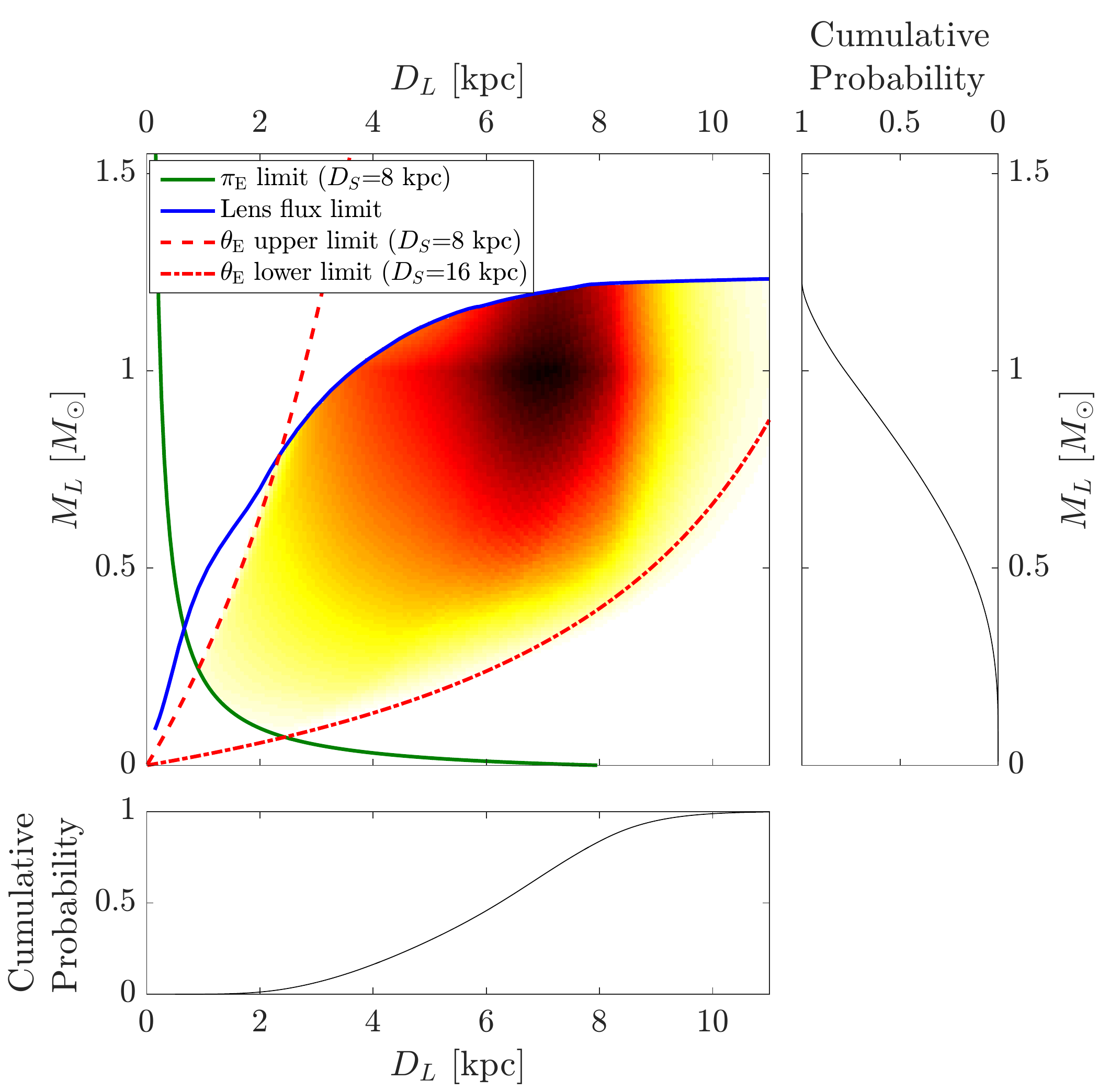}
\caption{Posterior probability distributions of the host mass and distance derived from the Bayesian analysis.
Also shown are the limits on the lens flux (blue), on $\theta_E$ for either bulge or far disk sources (red), and on the microlens parallax for a bulge source (green).
}
\label{fig:bayeisan}
\end{figure}

\begin{figure} 
\plotone{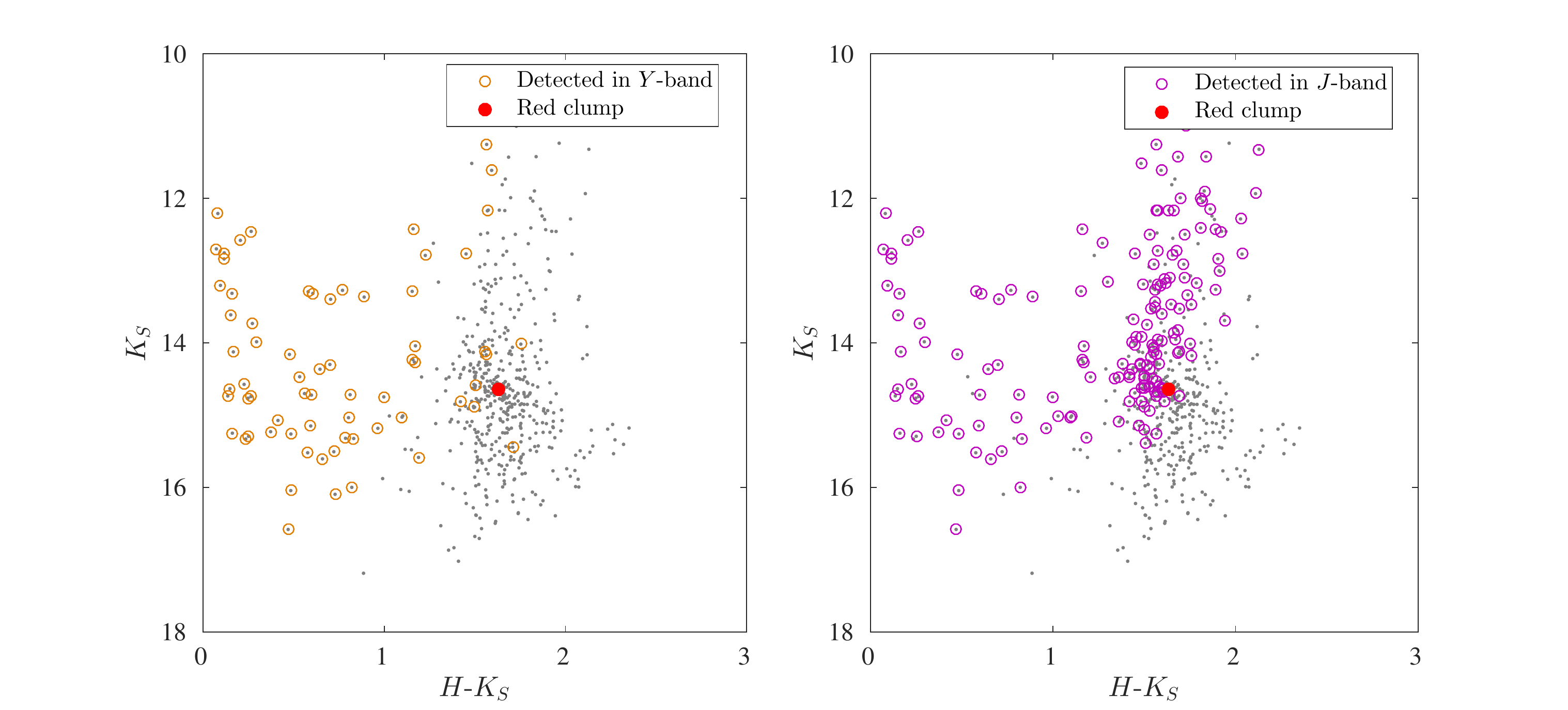}
\caption{Detection limits of DECam $Y$-band data (left) and VVV $J$-band data (right). These limits are brighter than expected using the extinction and reddening of the UKIRT $HK_S$ data (underlying gray points) and standard extinction curve, suggesting a non-standard and steeper extinction law.
}
\label{fig:ext_law}
\end{figure}

\end{document}